\documentclass{article}

\usepackage{microtype}
\usepackage{graphicx}
\usepackage{subfigure}
\usepackage{booktabs} 
\usepackage{eqnarray}
\usepackage{amsmath,amsthm}
\usepackage{amssymb}
\usepackage{mathtools}
\usepackage{multirow}
\usepackage{hyperref}
\usepackage{xcolor,colortbl}
\usepackage{color}

\usepackage[accepted]{icml2018}

\icmltitlerunning{Hawkes Process Kernel Structure Parametric Search with Renormalization Factors}

\begin{document}

\twocolumn[
\icmltitle{Hawkes Process Kernel Structure Parametric Search \\ with Renormalization Factors}



\icmlsetsymbol{equal}{*}

\begin{icmlauthorlist}
\icmlauthor{Rafael Lima}{unist}
\icmlauthor{Jaesik Choi}{unist}
\end{icmlauthorlist}

\icmlaffiliation{unist}{Department of Computer Engineering, Ulsan National Institute of Science and Technology, Ulsan, Republic of Korea}
\icmlcorrespondingauthor{Jaesik Choi}{jaesik@unist.ac.kr}

\icmlkeywords{Hawkes Processes, Automatic Report Generation, Kernel Search}

\vskip 0.3in
]



\printAffiliationsAndNotice{\icmlEqualContribution} 

\begin{abstract}
Hawkes Processes are a type of point process for modeling self-excitation, i.e., when the occurrence of an event makes future events more likely to occur. The corresponding self-triggering function of this type of process may be inferred through an Unconstrained Optimization-based method for maximization of its corresponding Loglikelihood function. Unfortunately, the non-convexity of this procedure, along with the ill-conditioning of the initialization of the self-triggering function parameters, may lead to a consequent instability of this method. Here, we introduce  Renormalization Factors, over four types of parametric kernels, as a solution to this instability. These factors are derived for each of the self-triggering function parameters, and also for more than one parameter considered jointly. Experimental results show that the Maximum Likelihood Estimation method shows improved performance with Renormalization Factors over sets of sequences of several different lengths.
\end{abstract}

\section{Introduction}

Point Processes \cite{VJ03} are mathematical objects for counting occurrences of discrete events over a continuous space. They have long been used for modeling a myriad of real-world phenomena, such as war deaths \cite{EL11}, social network interactions \cite{QZ15}, earthquake occurrences \cite{YO99} academic papers' citation counts \cite{HZ16} and stock orders \cite{EB15}.

In the case of the space considered being a one-dimensional real-valued axis, we may associate its coordinate to a time value, and deal with a so-called `purely temporal point process'. The simplest case of point process, the `Homogeneus Poisson Process', assumes that the arrival of events follow a static-valued rate, the so-called `Intensity Function', and that the events are independent of each other. However, it has been shown \cite{YL16,HX16} that this independence assumption is insufficient for modeling situations in which the data presents infectiousness or time-clustering effects, which indeed point to some causality mechanism among consecutive events.

Such temporally clustered behavior, also referred to as `Self-Excitation', can be modeled through a type of point process entitled `Hawkes Process' (HP), which models this behavior through the addition of an extra term in the Intensity Function of the process, which includes the influence of previous event occurrences in the current value of the arrival rate.

Most of the research concerning HPs consists of modeling/identifying the self-exciting term of the intensity function, hereby referred to as `Self-Exciting Kernel' (SEK), from one or more realizations of the process. One of the proposed methods for such inference consists of maximizing the corresponding parametric expression of the loglikelihood given a family of possible functions, such as exponential or power-law \cite{YO99,TO79}.

Unfortunately, the non-convexity of this procedure \cite{YL18}, along with the ill-conditioning of the initialization of the self-triggering function parameters, may lead to a consequent instability of this method. This means that the final parameter values found by the method correspond to a self-exciting kernel which models a corresponding process in which, in the long term, lead to an infinite amount of events occurring in a finite time interval. This happens when the parameters of the self-exciting function do not satisfy the so-called `stationarity criterion'.

Recent works \cite{EB12,TJ15} point to HPs parameter values near unstable configurations as good fitting for some types of data. However, the former just discuss the normalization over the amplitude parameter of exponential kernel, while the second latter deals solely with asymptotic properties, i.e., when the length of observation of the process tends to infinity ($T \rightarrow \infty$). Theoretical analysis of the stability of HPs is also done in \cite{DK13}.

In this work, we:
\begin{enumerate}
\item Describe the Unconstrained Optimization-based Hawkes Process Loglikelihood maximization procedure for four parametric families of functions (Exponential, Power-Law, Rayleigh and Tsallis Q-Exponential), deriving the required closed-form mathematical expressions;
\item Propose Renormalization Factors as instruments for constraining the Parametric MLE procedure to stable configurations of SEK parameters, which also take advantage of the previously discussed near-instability regions;
\item Derive closed-form expressions for the Renormalization Factors of four parametric families (Exponential, Power-Law, Rayleigh, Tsallis Q-Exponential) and show the several possible ways through which this stabilization can be done, with these factors being calculated over each parameter and also over more than one parameter considered jointly;
\item Validate the efficacy of the proposed method through experiments.
\end{enumerate}

\section{Related Work}

Regarding the the utilization of parametric functions for modeling SEKs in HPs: \cite{JE16} uses exponential kernels for modeling quick-decay in finance or web data. \cite{YO99} models slow decay influence with power-law kernels in earthquake, while \cite{QZ15} performs power law modeling experiments with social media-related data.

More recent works, s.a. the neural network-based  Hawkes processes in \cite{ND16,HM17} and the time-dependent Hawkes process (TiDeH) \cite{RK16}, allow for learning very flexible Hawkes processes with highly complicated intensity functions, while depending on the size and the quality of data.

\section{Hawkes Processes}

In this section, we will briefly discuss the theoretical concepts and mathematical definitions concerning the HP-based modeling of self-excitation or self-triggering behavior in Time Series data.

A realization of an one-dimensional point process, composed by a sequence of N time-events, can be expressed by a vector of the form ($t_{1}$,$t_{2}$, ... , $t_{n}$). Considering the real line as a time coordinate axis, it is possible to associate such vector with a so-called \textit{Counting Process} $N(t)$, with its derivative, $dN(t)$, which is set to 1 when there is an event at time t, and 0, otherwise. 

Moreover, such a point process is associated with a corresponding \textit{Intensity Function} $\lambda (t)$, which corresponds to the instantaneous expected rate of arrival of events, or the expectation of derivative of the counting process $N(t)$, i.e.:
\begin{eqnarray}
\lambda (t) = \lim_{h \to 0} \frac{\mathbb{E} [N(t+h)-N(t)]}{h}
\label{eq: intenshp}
\end{eqnarray}
The simplest example of this function would be a constant-valued arrival rate, $\mu$, as in the case of the \textit{Homogeneous Poisson Process} (HPP).

As previously stated, HPs model the intensity function in terms of \textit{self-excitation}: when the arrival of an event makes subsequent arrivals more likely to happen \cite{PL15};
This type of behavior can be described through an Intensity Function expression which takes into account the past events into its current value, i.e., a Conditional Intensity Function $\lambda (t)$, which may be described as such:
\begin{eqnarray*}
\lim_{h \to 0} \dfrac{\mathbb{E} (N(t + h) - N(t)| \mathcal{H} (t))}{h} \notag = \mu + \int_{-\infty}^{t} \phi (t-u) dN(u),
\end{eqnarray*}
where $\mathcal{H} (t)$ is the \textit{History} of the process, the set containing all the events up to time t; $\mu$ is denominated \textit{background rate}, or \textit{exogenous intensity}, and here is considered constant, such as the mean of a HPP; and $\phi (t)$ is referred to as \textit{self-exciting kernel}, or \textit{excitation function}.
From the definition in \cite{AH71}, we have that, given:
\begin{eqnarray}
||\phi|| = \int_{0}^{\infty} \phi (t) dt \leq 1,
\end{eqnarray}
then the corresponding process will attain wide-sense stationary behavior, from which the asymptotic steady arrival rate, also referred to as `first-order statistics', 
\begin{eqnarray}
\Lambda = \tfrac{\mu}{(1-||\phi||)},
\end{eqnarray}
can be evaluated, along with its stationary temporal covariance function, or `second-order statistics', which is independent of t:
\begin{eqnarray}
\nu (\tau) =  \mathbb{E} (dN(t) dN(t + \tau)).
\end{eqnarray}
The estimation of both $\Lambda$ and $\nu (\tau)$ draws on wide-sense stationarity assumptions which, apart from analytical convenience, are also related to the fact that, when dealing with real data and practical applications, the chain of causally triggered events will always be considered finite.

\section{Parametric Kernels for Hawkes Processes}

A concise way of expressing the time-clustering influence term $\phi$ is through the use of simply defined parametric functions. In this section, we describe the four types of function used in the present work, along with their main theoretical and practical usage motivation.

\textbf{Exponential (EXP($\alpha$,$\beta$))}. The exponential function is hereby defined as:
\begin{equation}
EXP(\alpha,\beta) = \alpha e^{-\beta t}
\end{equation}
It has been widely adopted in HP inference tasks, since it represents a general type of decay together with some analytical conveniences, s.a., the markovian property which allows the intensity at a given event coordinate to be calculated directly from the previous event, without the need of scanning throughout all the previous events.

\textbf{Power-Law (PWL(K,c,p))}. The power-law function is hereby defined as:
\begin{equation}
PWL(K,c,p) = \dfrac{K}{(t+c)^p}
\end{equation}
It has been used for modeling a slower form of decay than the exponential.

\textbf{Tsallis Q-Exponential (QEXP(a,q))}. The Tsallis Q-Exponential is hereby defined as:
\begin{equation}
QEXP(a,q) = \\ \nonumber
\end{equation}
\begin{equation}
\left\{
\begin{array}{ll}
a e^{-t} & \quad q = 1 \\
a \left[ 1+ (q-1)t\right]^\frac{1}{(1-q)} & \quad q \neq 0 \text{ and } 1+ (1-q)t > 0 \\
0 & q \neq 0 \text{ and } 1+ (1-q)t\leq 0
\end{array}
\right.
\end{equation}
It is a power transform, in other words, a continuous deformation , along the shape parameter `q' among Power-Law and Exponential-like triggering behaviours. Widely used in quantum optics and atomic physics, in this work, it is proposed for modeling decay in a more hybrid way between exponential and power-law. Since it has been shown that the behaviour of power-laws and exponentials may be very similar in some situations \cite{TB06}, a kernel such as Tsallis Q-Exponential may handle this in a unified way.

It is also widely used in statistics for variance stabilization, for turning the data more similar to some distribution, and also for improvement of the validity of association measures among variables (e.g. Pearson correlation).

\textbf{Rayleigh (RAY($\gamma$,$\eta$))}. The rayleigh kernel is hereby defined as:
\begin{equation}
RAY(\gamma,\eta) = \gamma t e^{-\eta t^2}
\end{equation}
It has been used in the context of survival times over diffusion networks \cite{MG13} for modeling a distinct, non-monotonically decaying, type of influence. It has been shown that some types of data, such as Stack Overflow response times, are not well suited to strictly decaying functions, exponential or power-law \cite{ND16}, and possess a more skewed-like distribution shape.

\section{Maximum Likelihood Estimation for Parametric SEK}

In this section, we describe the Unconstrained Optimization-based MLE procedure for HPs, and also derive closed-form expressions for the loglikelihood of four proposed families of parametric kernels. It was initially proposed in \cite{TO79}, for exponential functions, and a great deal of follow-up work developed it for other types of continuous and smooth functions. 
Given a sequence of time events $(t_1,t_2,...,t_n)$ and a defined intensity function $\lambda(t)$, its corresponding loglikelihood (llh) is given by:
\begin{eqnarray}
llh = \sum_{i=1}^{k} \log (\lambda (t_i)) - \int_{0}^{T} \lambda (u) du
\label{LLH}.
\end{eqnarray}
Since the intensity function of a HP possesses a closed-form expression given by Equation \ref{eq: intenshp}, we can calculate closed-form expressions of llh for each of the parametric kernels introduced in the previous section:
\subsection{EXP}
\begin{multline}
llh(EXP(\alpha,\beta)) = \sum_{i} \log (\mu + \sum_{j < i} \alpha e^{t_i - t_j}) \\- (\mu T + \sum_{i} \dfrac{\alpha}{\beta} (1 - e^{-\beta T-t_i}))
\end{multline}
\subsection{PWL}
\begin{multline}
llh(PWL(K,c,p)) = \sum_i \log(\mu + \sum_{j<i} \dfrac{K}{(t_i-t_j+c)^p}) \\- (\mu T + \sum_i \dfrac{K}{(p-1)}(c^{(1-p)}-(T-t_i + c)^{(1-p)})
\end{multline}
\subsection{QEXP}
\begin{itemize}
\item $1<q<2$:
\begin{equation}
\begin{split}
&llh(QEXP(a,q)) = \sum_i \log (\mu \\ &+ \sum_{j<i} a[1+(q-1)(t_i-t_j)]^{\dfrac{1}{1-q}})) \\&- (\mu T + \sum_i \dfrac{a}{2-q} (1-[1+(q-1)(T-t_i)]^{\dfrac{2-q}{1-q}}))
\end{split}
\end{equation}
\end{itemize}
\subsection{RAY}
\begin{equation}
\begin{split}
&llh(RAY(\gamma,\eta)) = \sum_i \log\left(\mu+\sum_{j<i} e^{-\eta (t_j-t_i)^2}\right) \\&-(\mu T + \sum_i \dfrac{\gamma}{2 \eta} (1-e^{-\eta (T-t_i)^2}))
\end{split}
\end{equation}

The standard MLE procedure consists in, given said parametric expressions for the loglikelihood, finding the corresponding SEK function parameters which maximize it. This search is done through Unconstrained Optimization methods, such as Gradient Descent, L-BFGS, or Nelder-Mead, over one or more sequences of events corresponding to a given HP realization.

\section{Renormalization Factors}

In this section, we introduce our stabilization method, entitled Maximum Likelihood Estimation with Renormalization Factors (RF-MLE), for stabilizing the solution of the Unconstrained Optimization-based MLE procedure. Given an possibly unstable SEK parameter configuration, from each of the four proposed SEKs, i.e., a tuple of parameters for which the value of $|\phi|$ is greater than 1, we find new tuples of parameter values for which the new value $|\phi|$, here defined as $|\phi|_{R}$ is equivalent to:
\begin{equation}
|\phi|_{R} = \dfrac{1}{1+\epsilon},
\end{equation}
with the strictly positive parameter $\epsilon$, here taken as a `safety margin', chosen as $10^{-1}$,$10^{-2}$ and $10^{-3}$, for our experiments.

\begin{figure}
\centering
\includegraphics[width=0.9\linewidth]{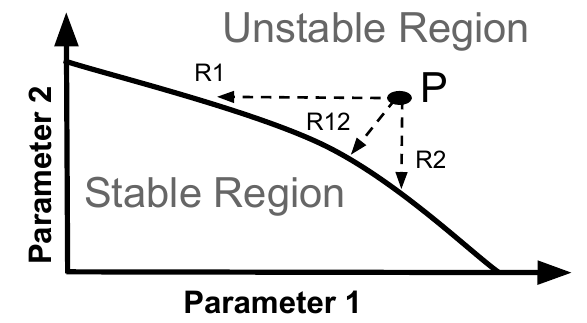}
\caption{Visual illustration of the Renormalization concept. Given original unstable parameters, at point P, we find stable configurations over Parameter 1 (R1), over Parameter 2 (R2) and jointly over both parameters (R12).}\label{fig: RF}
\end{figure}

These Renormalization Factors are derived for each parameter individually, and also for their joint combinations. This concept is illustrated in Figure \ref{fig: RF}.

\textbf{Exponential}. Given the EXP($\alpha$,$\beta$) function, the Renormalization Factors are found as in the following.\footnote{Here, $\hat{\Lambda}$ is calculated as $\dfrac{N}{T}$, with N being the total number of events contained in a given sequence.}
\begin{enumerate}
\item Renormalization over $\alpha$: 
\begin{equation}
(\mu',\alpha',\beta') = (\hat{\Lambda} \left( 1-1/(1+\epsilon)\right),\frac{\alpha}{|\phi| (1+\epsilon)},\beta)
\end{equation}
\item Renormalization over $\beta$: 
\begin{equation}
(\mu',\alpha',\beta') = (\hat{\Lambda} \left( 1-1/(1+\epsilon)\right),\alpha,\beta |\phi| (1+\epsilon))
\end{equation}
\item Joint Renormalization over $\alpha$ and $\beta$: 
\begin{equation}
\begin{split}
&(\mu',\alpha',\beta') = \\& (\hat{\Lambda} \left( 1-1/(1+\epsilon)\right),\frac{\alpha}{\sqrt{|\phi| (1+\epsilon)}},\beta\sqrt{|\phi|(1+\epsilon)})
\end{split}
\end{equation}
\end{enumerate}
\textbf{Power-Law}. Given the PWL(K,c,p) function, the Renormalization Factors are found as in the following.
\begin{enumerate}
\item Renormalization over K: 
\begin{equation}
(\mu',K',c',p') = (\hat{\Lambda} \left( 1-1/(1+\epsilon)\right),\frac{K}{|\phi|(1+\epsilon)},c,p)
\end{equation}
\item Renormalization over c: 
\begin{equation}
\begin{split}
&(\mu',K',c',p') = \\& (\hat{\Lambda} \left( 1-1/(1+\epsilon)\right),K,c(|\phi|(1+\epsilon))^{\frac{1}{p-1}},p)
\end{split}
\end{equation}
\item Renormalization over p: 
\begin{equation}
\begin{split}
&(\mu',K',c',p') = \\&(\hat{\Lambda} \left( 1-1/(1+\epsilon)\right),K,c,1+\frac{W(\Delta \log c)}{\log c})
\end{split}
\end{equation}
where 
\begin{equation}
\Delta = |\phi| (1+\epsilon) (p-1) c^{(p-1)}
\end{equation} 
and W is the analytical continuation of branch 0 of the product log (Lambert-W) function.
\item Joint Renormalization over K and c: 
\begin{equation}
\begin{split}
&(\mu',K',c',p') = \\&(\hat{\Lambda} \left( 1-1/(1+\epsilon)\right),\frac{K}{\sqrt{|\phi|(1+\epsilon)}},c(\sqrt{|\phi|(1+\epsilon)})^{\frac{1}{p-1}},p)
\end{split}
\end{equation}
\item Joint Renormalization over K and p: 
\begin{equation}
\begin{split}
&(\mu',K',c',p') = \\&(\hat{\Lambda} \left( 1-1/(1+\epsilon)\right),\frac{K}{\sqrt{|\phi|(1+\epsilon)}},c,1+\frac{W(\Delta \log c)}{\log c})
\end{split}
\end{equation}
where 
\begin{equation}
\Delta = \sqrt{|\phi| (1+\epsilon)} (p-1) c^{(p-1)}
\end{equation}
\end{enumerate}
\textbf{Tsallis Q-Exponential}. Given the QEXP(a,q) function, the Renormalization Factors are found as in the following.
\begin{enumerate}
\item Renormalization over a: 
\begin{equation}
(\mu',a',q') = (\hat{\Lambda} \left( 1-1/(1+\epsilon)\right),\frac{a}{|\phi|(1+\epsilon)},q)
\end{equation}
\item Renormalization over q: 
\begin{equation}
(\mu',a',q') = (\hat{\Lambda} \left( 1-1/(1+\epsilon)\right),a,2-(2-q)|\phi|(1+\epsilon))
\end{equation}
\item Joint Renormalization over a and q: 
\begin{equation}
\begin{split}
&(\mu',a',q') = (\hat{\Lambda} \left( 1-1/(1+\epsilon)\right),\\&\frac{a}{\sqrt{|\phi|(1+\epsilon)}},2-(2-q)\sqrt{|\phi|(1+\epsilon)})
\end{split}
\end{equation}
\end{enumerate}
\textbf{Rayleigh}. Given the RAY($\gamma$,$\eta$) function, the Renormalization Factors are found as in the following.
\begin{enumerate}
\item Renormalization over $\gamma$: 
\begin{equation}
(\mu',\gamma',\eta') = (\hat{\Lambda} \left( 1-1/(1+\epsilon)\right),\frac{\gamma}{|\phi| (1+\epsilon)},\eta)
\end{equation}
\item Renormalization over $\eta$: 
\begin{equation}
(\mu',\gamma',\eta') = (\hat{\Lambda} \left( 1-1/(1+\epsilon)\right),\gamma,\eta |\phi| (1+\epsilon))
\end{equation}
\item Joint Renormalization over $\gamma$ and $\eta$: 
\begin{equation}
\begin{split}
&(\mu',\gamma',\eta') = (\hat{\Lambda} \left( 1-1/(1+\epsilon)\right),\\&\frac{\gamma}{\sqrt{|\phi| (1+\epsilon)}},\eta\sqrt{|\phi|(1+\epsilon)})
\end{split}
\end{equation}
\end{enumerate}

\section{Experiments}

For the experiments, we were interested in how the types of renormalization affect the performance of the standard MLE algorithm across all the four types of kernels.
We trained the models on 10 sequences for T=1000,5000, 10000 and 30000, independently simulated with each of the four types of kernels, with the respective parameters: $\mu=0.5$, $\alpha=0.06$, $\beta=0.2$, $K=0.06$, $c=1.0$, $p=11$, $a=0.06$, $q=0.5$, $\gamma=0.06$ and $\eta=0.2$.

The optimization procedure was maximizing the Loglikelihood Functions for each of the four proposed SEKs over each realization of the sequences, using the Nelder-Mead method \cite{NM65}.

\begin{figure*}[ht!]
\centering     
\subfigure[Results for T=1000.]{\label{fig: T1000}\includegraphics[width=0.75\linewidth]{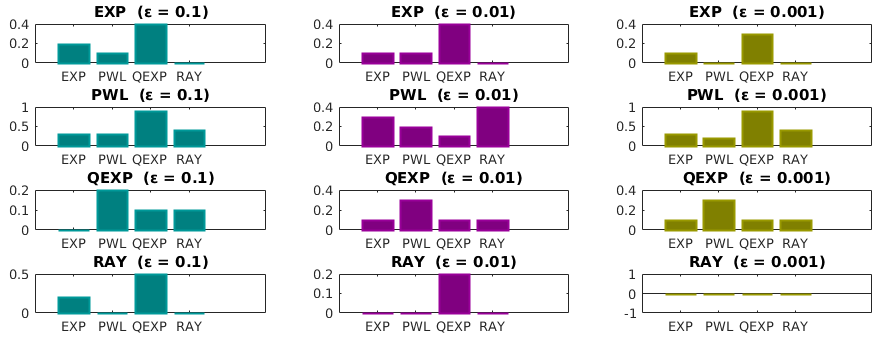}}
\subfigure[Results for T=5000.]{\label{fig: T5000}\includegraphics[width=0.75\linewidth]{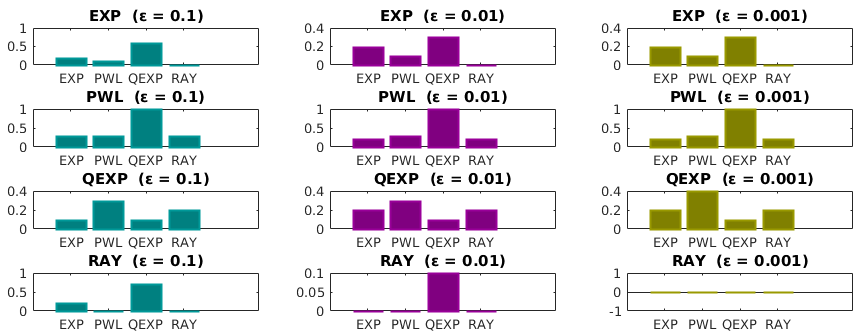}}
\subfigure[Results for T=10000.]{\label{fig: T10000}\includegraphics[width=0.75\linewidth]{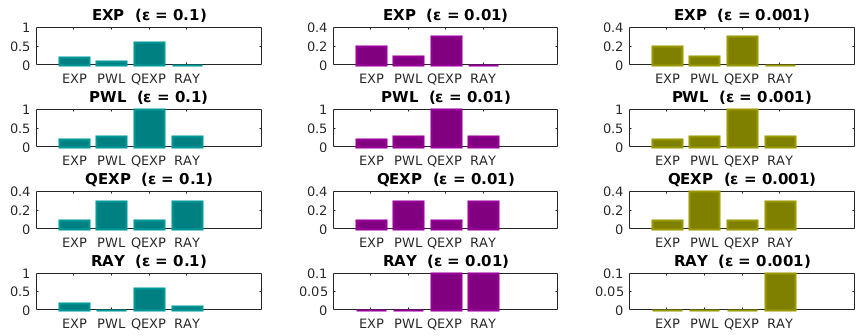}}
\subfigure[Results for T=30000.]{\label{fig: T30000}\includegraphics[width=0.75\linewidth]{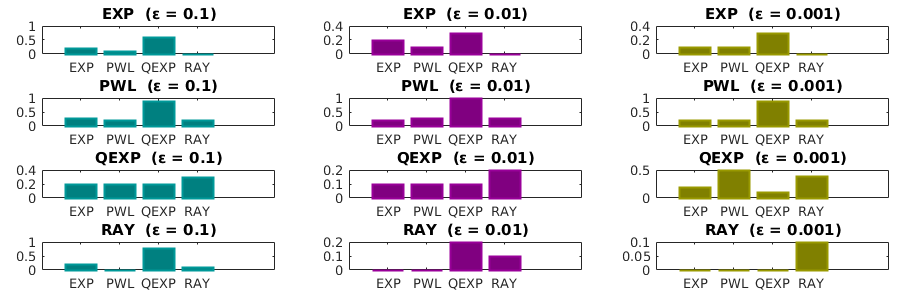}}
\caption{Ratio of sequences for which the RF-MLE showed improvement over the standard MLE on the sequences with T = 1000, 5000, 10000 and 30000. This ratio was calculated for each kernel type over their corresponding sequences and also in the case of misassignment of kernel, i.e., when a given kernel type is used on sequences simulated with other kernel types. For example, the upper leftmost barplot shows the ratio of sequences, over the total number of sequences simulated with each kernel type, for which the RF-MLE for the kernel EXP, with $\epsilon$ set as 0.1, showed an improvement of the average loglikelihood over the sequences, comparing with the standard MLE.}\label{fig: freqmle}
\end{figure*}

The goal was to verify the average difference among the loglikelihood from the search with RFs (RF-MLE), which outputs the highest loglikelihood value among the original MLE and all the possible renormalizations, and the original version of the MLE method. The results are shown in Figures \ref{fig: T1000}, \ref{fig: T5000}, \ref{fig: T10000} and \ref{fig: T30000}for sequences with T=1000, 5000, 10000 and 30000, respectively.

\begin{table*}[ht!]

\parbox{.49\linewidth}{
\resizebox{1.\linewidth}{!}{
\centering
\begin{tabular}{|c||c c c c|}
\hline
\multirow{3}{*}{\vbox{\hbox{\strut Sequence}\hbox{\strut Type}}} & \vbox{\hbox{\strut MLE}\hbox{\strut}}  & \vbox{\hbox{\strut RF-MLE}\hbox{\strut ($\epsilon = 0.1$)}}  & \vbox{\hbox{\strut RF-MLE}\hbox{\strut ($\epsilon = 0.01$)}}  & \vbox{\hbox{\strut RF-MLE}\hbox{\strut ($\epsilon = 0.001$)}}  \\
\cline{2-5}
 & \multicolumn{4}{|c|}{Selected Kernel}\\
\cline{2-5}
 & \multicolumn{4}{|c|}{EXP}\\
\hline

EXP & -933.35  & \textbf{-915.78} & -917.01 & -917.337 \\
PWL & -874.01  & \textbf{-869.30} & -873.28 & -874.01 \\
QEXP & -607.02 & \textbf{-576.74} & -583.42 & -584.68 \\
RAY & \textbf{-905.70} & \textbf{-905.70} & \textbf{-905.70} & \textbf{-905.70} \\
\hline
  & \multicolumn{4}{|c|}{PWL}\\
\hline
EXP & -1222.95 & \textbf{-1140.89} & -1203.14 & -1207.36 \\
PWL & -1662.69 & \textbf{-1550.44} & -1600.95 & -1604.40 \\
QEXP & -979.52 & \textbf{-701.06} & -740.48 & -745.00 \\
RAY & -1338.68 & \textbf{-1221.07} & -1221.72 & -1226.69 \\
\hline
  & \multicolumn{4}{|c|}{QEXP}\\
\hline
EXP & \textbf{-781.88} & \textbf{-781.88} & \textbf{-781.88} & \textbf{-781.88} \\
PWL & -848.25 & -848.25 & -848.25 & \textbf{-559.70} \\
QEXP & -779.39 & \textbf{-760.67} & \textbf{-760.67} & -779.39 \\
RAY & -850.53 & -823.90 & \textbf{-781.88} & -821.09 \\
\hline
  & \multicolumn{4}{|c|}{RAY}\\
\hline
EXP & -943.67 & \textbf{-934.11} & -943.67 & -943.67 \\
PWL & \textbf{-877.46} & \textbf{-877.46} & \textbf{-877.46} & \textbf{-877.46} \\
QEXP & -583.69 & \textbf{-566.26} & -581.10 & -583.69 \\
RAY & \textbf{-913.76} & \textbf{-913.76} & \textbf{-913.76} & \textbf{-913.76} \\
\hline

\hline
\end{tabular}

}
\caption{Performance comparison of MLE and RF-MLE, for sequences with T=5000.}
}
\hfill
\parbox{.49\linewidth}{
\resizebox{1.\linewidth}{!}{
\centering
\begin{tabular}{|c||c c c c|}
\hline
\multirow{3}{*}{\vbox{\hbox{\strut Sequence}\hbox{\strut Type}}} & \vbox{\hbox{\strut MLE}\hbox{\strut}}  & \vbox{\hbox{\strut RF-MLE}\hbox{\strut ($\epsilon = 0.1$)}}  & \vbox{\hbox{\strut RF-MLE}\hbox{\strut ($\epsilon = 0.01$)}}  & \vbox{\hbox{\strut RF-MLE}\hbox{\strut ($\epsilon = 0.001$)}}  \\
\cline{2-5}
 & \multicolumn{4}{|c|}{Selected Kernel}\\
\cline{2-5}
 & \multicolumn{4}{|c|}{EXP}\\
\hline
EXP & -4676.61  & \textbf{-4599.52} & -4605.21 & -4606.31 \\
PWL & -4409.35  & \textbf{-4386.74} & -4404.62 & -4407.94 \\
QEXP & -3133.68 & \textbf{-2998.18} & -3014.98 & -3017.27 \\
RAY & \textbf{-4479.72} & \textbf{-4479.72} & \textbf{-4479.72} & \textbf{-4479.72} \\
\hline
  & \multicolumn{4}{|c|}{PWL}\\
\hline
EXP & -6427.87 & -6049.23 & -6034.03 & \textbf{-6031.82}\\
PWL & -8732.92 & -8324.04 & -8294.48 & \textbf{-8291.94} \\
QEXP & -5162.15 & \textbf{-3550.52} & -3605.26 & -3612.18 \\
RAY & -6198.44 & -6031.68 & -6022.75 & \textbf{-6022.20}\\
\hline
  & \multicolumn{4}{|c|}{QEXP}\\
\hline
EXP & -4263.04 & \textbf{-3577.93} & -4249.27 & -4263.04 \\
PWL & -2246.73 & \textbf{-1623.26} & -2246.73 & -1994.41 \\
QEXP & -4834.63 & -\textbf{3824.03} & -4824.63 & -4834.63 \\
RAY & \textbf{-1898.07} & \textbf{-1898.07} & \textbf{-1898.07} & \textbf{-1898.07} \\
\hline
  & \multicolumn{4}{|c|}{RAY}\\
\hline
EXP & \textbf{-4726.30} & \textbf{-4726.30} & \textbf{-4726.30} & \textbf{-4726.30} \\
PWL & \textbf{-4439.19} & \textbf{-4439.19} & \textbf{-4439.19} & \textbf{-4439.19} \\
QEXP & -2991.98 & \textbf{-2931.48} & -2985.19 & -2991.98 \\
RAY & \textbf{-4507.61} & \textbf{-4507.61} & \textbf{-4507.61} & \textbf{-4507.61} \\
\hline

\hline
\end{tabular}
}
\caption{Performance comparison of MLE and RF-MLE, for sequences with T=5000.}
}
\centering
\parbox{.49\linewidth}{
\resizebox{1.\linewidth}{!}{
\centering
\begin{tabular}{|c||c c c c|}
\hline
\multirow{3}{*}{\vbox{\hbox{\strut Sequence}\hbox{\strut Type}}} & \vbox{\hbox{\strut MLE}\hbox{\strut}}  & \vbox{\hbox{\strut RF-MLE}\hbox{\strut ($\epsilon = 0.1$)}}  & \vbox{\hbox{\strut RF-MLE}\hbox{\strut ($\epsilon = 0.01$)}}  & \vbox{\hbox{\strut RF-MLE}\hbox{\strut ($\epsilon = 0.001$)}}  \\
\cline{2-5}
 & \multicolumn{4}{|c|}{Selected Kernel}\\
\cline{2-5}
 & \multicolumn{4}{|c|}{EXP}\\
\hline
EXP & -9345.30  & \textbf{-9184.65} & -9195.60 & -9197.34 \\
PWL & -8804.21  & \textbf{-8758.80} & -8795.91 & -8803.54 \\
QEXP & -6224.90 & \textbf{-5924.63} & -5959.67 & -5963.72 \\
RAY & \textbf{-9011.67} & \textbf{-9011.67} & \textbf{-9011.67} & \textbf{-9011.67} \\
\hline
  & \multicolumn{4}{|c|}{PWL}\\
\hline
EXP & -13224.13 & -12475.77 & -12884.69 & \textbf{-12449.9} \\
PWL & -18001.13 & -17281.74 & -17677.44 & \textbf{-17214.11} \\
QEXP & -9909.25 & \textbf{-7069.38} & -7159.89 & -7168.51 \\
RAY & -12653.81 & -12554.38 & -12653.81 & \textbf{-12526.31} \\
\hline
  & \multicolumn{4}{|c|}{QEXP}\\
\hline
EXP & -9364.55 & \textbf{-7954.48} & -9364.55 & -9364.55 \\
PWL & -4228.78 & -4153.88 & -4228.78 & \textbf{-3465.89} \\
QEXP & -9638.70 & \textbf{-7629.54} & -9638.70 & -9638.70 \\
RAY & -3625.10 & -3555.13 & -3541.83 & \textbf{-3530.86} \\
\hline
  & \multicolumn{4}{|c|}{RAY}\\
\hline
EXP & -9430.41 & \textbf{-9343.28} & -9430.41 & -9430.41 \\
PWL & \textbf{-8860.35} & \textbf{-8860.35} & \textbf{-8860.35} & \textbf{-8860.35} \\
QEXP & -5904.64 & \textbf{-5780.61} & -5894.93 & -5904.64 \\
RAY & -9119.86 & \textbf{-9074.78} & -9080.00 & -9080.81 \\
\hline

\hline
\end{tabular}
}
\caption{Performance comparison of MLE and RF-MLE, for sequences with T=10000.}
}
\parbox{.49\linewidth}{
\resizebox{1.\linewidth}{!}{
\centering
\begin{tabular}{|c||c c c c|}
\hline
\multirow{3}{*}{\vbox{\hbox{\strut Sequence}\hbox{\strut Type}}} & \vbox{\hbox{\strut MLE}\hbox{\strut}}  & \vbox{\hbox{\strut RF-MLE}\hbox{\strut ($\epsilon = 0.1$)}}  & \vbox{\hbox{\strut RF-MLE}\hbox{\strut ($\epsilon = 0.01$)}}  & \vbox{\hbox{\strut RF-MLE}\hbox{\strut ($\epsilon = 0.001$)}}  \\
\cline{2-5}
 & \multicolumn{4}{|c|}{Selected Kernel}\\
\cline{2-5}
 & \multicolumn{4}{|c|}{EXP}\\
\hline
EXP & -27977.34 & \textbf{-27478.56} & -27505.80 & -27509.08 \\
PWL & -26409.90 & \textbf{-26275.62} & -26384.06 & -26404.55 \\
QEXP & -18709.47 & \textbf{-17845.99} & -17966.39 & -17977.52 \\
RAY & \textbf{-27040.52} & \textbf{-27040.52} & \textbf{-27040.52} & \textbf{-27040.52} \\
\hline
  & \multicolumn{4}{|c|}{PWL}\\
\hline
EXP & -38778.95 & -37577.37 & -37474.50 & \textbf{-37463.87} \\
PWL & -55402.71 & -54297.97 & -54198.61 & \textbf{-54188.42} \\
QEXP & -29858.67 & \textbf{-21510.88} & -21701.71 & -21711.34 \\
RAY & -39847.99 & \textbf{-39310.19 }& -39332.57 & -39335.31 \\
\hline
  & \multicolumn{4}{|c|}{QEXP}\\
\hline
EXP & -25912.10 & \textbf{-21605.03} & -25912.10 & -25912.10 \\
PWL & -12480.82 & -12048.36 & -12480.82 & \textbf{-8821.05} \\
QEXP & -28863.15 & \textbf{-22992.06} & -28863.15 & -28863.15 \\
RAY & -12296.04 & -12011.74 & -12296.04 & -12296.04 \\
\hline
  & \multicolumn{4}{|c|}{RAY}\\
\hline
EXP & -28227.41 & \textbf{-27966.14} & -28227.41 & -28227.41 \\
PWL & \textbf{-26575.83} & \textbf{-26575.83} & \textbf{-26575.83} & \textbf{-26575.83} \\
QEXP & -17842.22 & \textbf{-17412.77} & -17810.90 & -17842.22 \\
RAY & -27328.09 & \textbf{-27200.80} & -27214.74 & -27216.62 \\
\hline

\hline
\end{tabular}
}
\caption{Performance comparison of MLE and RF-MLE, for sequences with T=30000.}
}
\end{table*}
It is possible to see that the RF-MLE shows significant improvement, over the standard MLE, on the Loglikelihood of the estimated model even when the kernel being fitted is misassigned, i.e., when a type of kernel is used to model a sequence generated by another type of kernel.

In Figure \ref{fig: freqmle}, it is possible to see the ratio of sequences, for each kernel and sequence types, in which the RF-MLE improves the final resulting loglikelihood.

\section{Conclusion}

Hawkes Processes are a type of point process for modeling self-excitation, i.e., when the occurrence of an event makes future events more likely to occur. We introduced  Renormalization Factors, over four types of parametric kernels, as a solution to the instability of the Unconstrained Optimization-based method for maximization of the Loglikelihood function of Hawkes Processes. These factors were derived for each of the self-triggering function parameters, and also for more than one parameter considered jointly. Experimental results demonstrated that the Parametric MLE estimation method shows improved performance with Renormalization Factors over sets of sequences of several different lengths.

\subsection{Future Work}

For further investigation, we plan to study the renormalization procedures over sums of each of the proposed parametric kernels, which will be done in a weighted, proportional way. We also plan to propose Renormalization Factors for the Multivariate HPs case.








\newpage

\section*{Acknowledgements}

The authors would like to thank Marcos Cleison Silva Santana, from Recogna/UNESP, for the fruitful discussions regarding the Tsallis Q-Exponential.






\nocite{langley00}

\bibliography{example_paper}
\bibliographystyle{icml2018}


\end{document}